# Optical absorption sensing with dual-spectrum silicon LEDs in SOI-CMOS technology

Satadal Dutta[1*], Peter G. Steeneken[1], and Gerard J. Verbiest[1]

[1]Precision and Microsystems Engineering, Delft University of Technology, Delft, The Netherlands
*Email: s.dutta-1@tudelft.nl

*Abstract*—Silicon p-n junction diodes emit low-intensity, broad-spectrum light near 1120 nm in forward bias and between 400-900 nm in reverse bias (avalanche). For the first time, we experimentally achieve optical absorption sensing of pigment in solution with silicon micro LEDs designed in a standard silicon-on-insulator CMOS technology. By driving a single LED in both forward and avalanche modes of operation, we steer it's electroluminescent spectrum between visible and near-infrared (NIR). We then characterize the vertical optical transmission of both visible and NIR light from the LED through the same micro-droplet specimen to a vertically mounted discrete silicon photodiode. The effective absorption coefficient of carmine solution in glycerol at varying concentrations were extracted from the color ratio in optical coupling. By computing the LED-specific molar absorption coefficient of carmine, we estimate the concentration (~0.040 mol L$^{-1}$) and validate the same with a commercial spectrophotometer (~0.030 mol L$^{-1}$). With a maximum observed sensitivity of ~1260 cm$^{-1}$mol$^{-1}$L, the sensor is a significant step forward towards low-cost CMOS-integrated optical sensors with silicon LED as the light source intended for biochemical analyses in food sector and plant/human health.

## I. Introduction

Silicon (Si) photonics is emerging as a key player in the development of CMOS-integrated optical devices for applications in bio-chemical sensing and data communication links [1]-[10]. State-of-the-art optical sensors, popular in biochemical analyses in both medical and food sector, use expensive lasers or quasi-monochromatic LEDs made of III-V compound semiconductors [11]. This prevents the monolithic integration with driver/read-out electronics designed in Si CMOS technology. Interestingly, Si p-n junction diodes exhibit broad-spectrum electroluminescence (EL) near 1120 nm in forward mode (FM) and in the range of 400 nm - 900 nm in avalanche mode (AM) of operation, although at a very low quantum efficiency (~10$^{-3}$-10$^{-5}$) [12]-[17] due to the indirect bandgap of Si. Recent advancements [18]-[20] have successfully highlighted the Si LED as a promising candidate for monolithically integrated optical interconnects due to the high responsivity of Si photodiodes (PDs) for wavelengths ($\lambda$) < 1000 nm.

In this work, we experimentally show for the first time that Si LEDs designed in a standard SOI-CMOS technology are viable for optical absorption sensing by driving a single LED in both FM and AM operation. The vertically transmitted light propagates through a pigmented micro-droplet placed on the surface followed by detection by a discrete Si PD mounted externally above the chip (see Fig. 1). From the colour ratio (AM to FM) of optical coupling to the Si PD and the mean height of the droplet (optical path length), we obtain the effective absorption coefficient ($\alpha$) of carmine solution specific to our broadband Si LED. The ability to electrically switch between visible and near-infrared emission from a single Si LED eliminates the need for any process modification or device replacement in the optical sensor. The 400 nm-1300 nm spectral range is highly suitable in bio-sensing e.g. photosynthetic pigments [23], leaf-water status [24][25], coloured contaminants in water [26]-[28] and blood oxygen levels [29]. Our work, therefore, constitutes a major step forward in realizing low-cost, and micro-volume CMOS-integrated optical sensors with silicon light source.

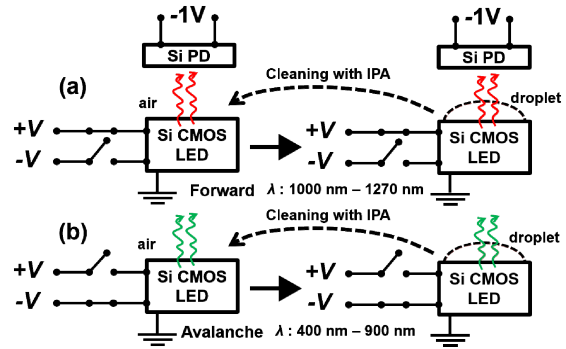

Fig. 1. Schematic block diagram illustrating the optical sensing method. Optical coupling from the Si LED on the CMOS chip to an externally mounted Si PD (reverse biased at 1 V) is measured in air and in presence of the same glycerol droplet specimen in both (a) forward and (b) avalanche modes of LED operation. The droplet contains dissolved carmine pigment absorbing light of $\lambda$ in the 400 nm - 600 nm interval, emitted by the LED in avalanche mode. Post-measurement of each droplet, the chip surface is cleaned with laboratory grade iso-propanol (IPA) and to re-use the LED for the next droplet measurement.

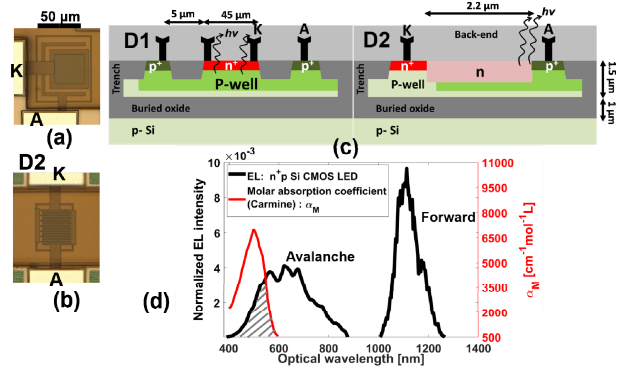

Fig. 2. Top-view micrograph of (a) n$^+$p junction LED (D1), (b) p$^+$nn$^+$ junction LED (D2) with their respective device cross-sections in a SOI CMOS technology [21] shown in (c). (d) Avalanche-mode (400 nm – 900 nm) and forward-mode (1000 nm – 1270 nm) normalized electroluminescent intensity spectra [18] of diode D1 (in black) measured with a ADC-1000-USB and AvaSpec-UV/VIS/NIR spectrometers respectively from Avantes B.V. The molar absorption coefficient spectrum (in red) of carmine [22] shows the overlap (grey stripes) within 400 nm – 600 nm AM EL spectrum of Si LED.

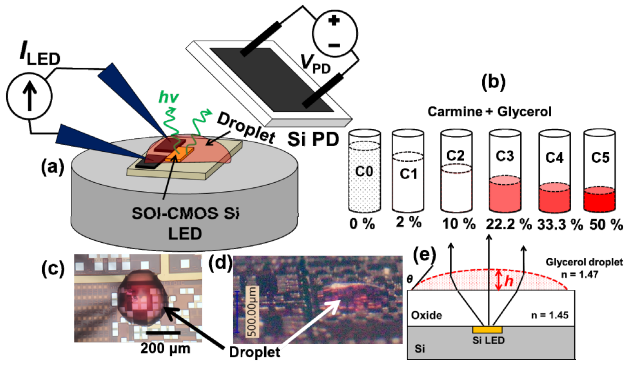
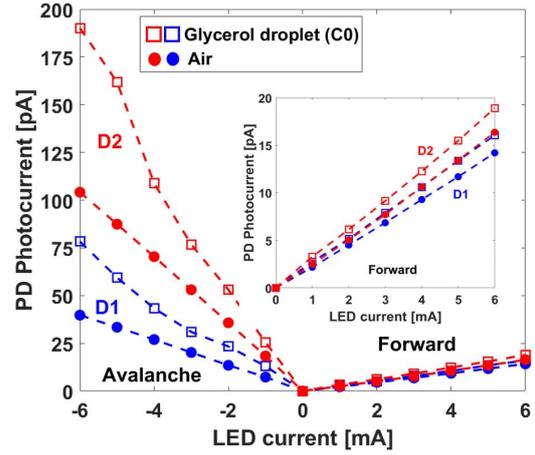

Fig. 3. (a) Schematic of the measurement set-up (not to scale). The on-chip Si LED and the external Si photodiode (BPW34) are driven by a 2-channel Keysight B2912A precision source-measure unit (SMU) in constant current mode and constant voltage mode respectively. (b) Solution specimens of carmine in laboratory-grade glycerol with the indicated concentrations (% by volume) relative to a reference commercial sample of liquid food color (camine (E120) in glycerol (E422) and water). (c) Top-view and (d) slanted view (25 degrees w.r.t. horizontal) of a droplet (from sample $c_2$). (e) Schematic cross-section showing the light-rays being focussed by the plano-convex micro lens formed by the droplet.

## II. EXPERIMENTAL MATERIALS AND METHOD

### A. LED design and electroluminescent spectra

Figs. 2a, 2b show the top-view layout of the two test LEDs D1 and D2 respectively, designed in a standard 130 nm silicon-on-insulator (SOI) CMOS technology [21]. Figs. 2c shows their vertical cross-section. D1 is a vertical $n^+p$ junction at a depth of ~0.25 μm with a breakdown voltage of ~17 V [18] and peripherally placed electrode contacts. D2 is a lateral $p^+nn^+$ junction reaching the Si-SiO2 interface, with an avalanche breakdown voltage of ~15 V. D2 has an interdigitated (comb-like) layout [30] of alternating cathode and anode fingers to yield a higher junction area and therefore a higher spatial uniformity and quantum efficiency ($\eta_{LED}$) of light emission. Fig. 2d shows the normalized EL-spectrum [18] of D1 in AM and FM modes of operation, alongside the literature reported [22] molar absorption coefficient ($\alpha_M$) of carmine that overlaps with the AM-EL spectrum for 400 nm < λ < 600 nm.

### B. Measurement set-up and sample preparation

As illustrated in Fig. 3a, the on-chip Si LED, placed on a vacuum chuck, is electrically probed by Tungsten needles. The Si PD (BPW34 from Vishay semiconductors), is mounted vertically above the chip at a centre-height ~ 5 mm tilted at ~ 45° via a micro-manipulator. The LED is driven in a constant current (sweep) mode and the PD is driven at a fixed reverse bias of 1V using a Keysight B2912A precision SMU with dc offset currents < 1 pA. Commercial carmine (E120 food colour) solution (of unknown concentration $c_{ref}$) is used as the target specimen which is diluted in glycerol solvent to yield samples of concentrations $c_1$=2(±0.4)%, $c_2$=10(±2)%, $c_3$=22.2(±4)%, $c_4$=33.3(±6)% and $c_5$=50(±5)% by volume relative to $c_{ref}$ (see Fig. 3b). Sample $c_0$=0 refers to only glycerol (no pigment). PD photocurrent ($\Delta I_{PD}$) is measured in air and in presence of a micro-droplet (diameter: 250 μm – 500 μm), which is transferred from each solution sample to the chip with a hydrophilic tip of a ~100 μm silica fiber (Figs. 3c,d) masking the Si LED entirely. The same droplet is used to measure $\Delta I_{PD}$ for AM and FM LED.

Fig. 4 Photocurrent (at reverse bias of 1 V) in the Si PD versus LED current in avalanche (- X axis) and forward modes (+X axis) of operation, in the presence (squares) and absence (circles) of a glycerol droplet (without carmine). (inset): Zoom-in of the forward mode of operation.

## III. RESULTS AND ANALYSIS

The glycerol droplet height $h$ is primarily governed by the angle of contact [31][32] at the local liquid - chip (SiO2) interface, and was estimated to be within 180 μm - 280 μm (Fig. 3c) by imaging the chip surface at an inclination of ~25° with a VHX microscope from Keyence. The same was verified by droplet imaging with a Attension Theta Lite optical tensiometer by experimenting with multiple droplet samples on the chip surface. A droplet acts as a microscopic plano-convex lens (Fig. 3e) with a good matching of refractive index (~1.47) with that of the back-end oxide layer (~1.45), which enhances the vertical transmission coefficient of light. Optical gain due to lensing is evident from Fig. 4 which shows the measured $\Delta I_{PD}$ versus $I_{LED}$ in AM and FM operation. For both D1 and D2, $\Delta I_{PD}^{AM} > \Delta I_{PD}^{AM}$ primarily due to the higher PD quantum efficiency ($\eta_{PD}$) for light emitted in AM [18]. $\Delta I_{PD}^{FM}$ and $\Delta I_{PD}^{AM}$ are respectively ~1.2 times and ~1.5 times higher in presence of a glycerol ($c_0$) droplet as compared to in air. A mismatch in the gain in AM and FM is likely due to electrostatic effects on glycerol refractive index [33] at different $V_{LED}$ applied in AM (~19 V) and FM (~1 V) operation. Further, at a given $I_{LED}$, both $\Delta I_{PD}^{FM}$ and $\Delta I_{PD}^{AM}$ is higher for D2 than for D1 due to the higher external quantum efficiency of D2. Since the pigment absorption window overlaps only with the AM EL spectrum, we can express the photocurrent in presence of the droplet as

$$\Delta I_{PD(Gly)}^{FM}(c_i) = \eta_{LED}^{FM} \cdot \eta_{PD}^{FM} \cdot \eta_{Gly}^{FM} \cdot I_{LED} \quad (1)$$

$$\Delta I_{PD(Gly)}^{AM}(c_i) = \eta_{LED}^{AM} \cdot \eta_{PD}^{AM} \cdot \eta_{Gly}^{AM} \cdot I_{LED} \cdot \exp\{-\alpha(c_i) \cdot h\} \quad (2)$$

Here $\eta_{Gly}$ is the light-extraction efficiency from the chip through the droplet. The colour ratio (*COR*) of optical coupling in AM to FM can then be expressed as:

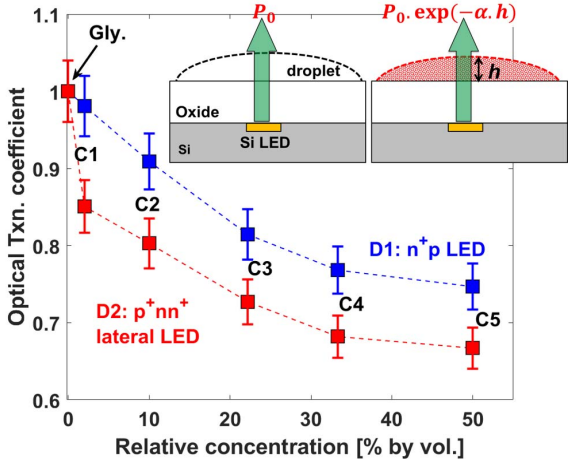

Fig. 5 Optical transmission coefficient [$\exp(-\alpha(c_i).h)$] through the micro-droplet with varying concentrations (relative to $c_{ref}$) of the carmine solution when illuminated by the CMOS LEDs D1 (blue) and D2 (red) and the optical intensity measured vertically with the external Si PD. The values are normalized to the case when only glycerol is present, where the transmission coefficient is assigned a value of 1 (i.e. $\alpha = 0$).

$$COR(c_i) = \{\Delta I_{PD(Gly)}^{AM}(c_i)/ \Delta I_{PD(Gly)}^{FM}(c_i)\}$$
$$= \{\Delta I_{PD(Gly)}^{AM}(c_0)/ \Delta I_{PD(Gly)}^{FM}(c_0)\} \cdot \exp\{-\alpha(c_i).h\} \quad (3)$$

Fig. 5 shows that the transmission coefficient ($T(c_i)=\exp(-\alpha(c_i).h)$), extracted at $I_{LED}$ =4 mA for D1 and D2, decreases with increasing $c_i$. The leading factor in the RHS of (3) i.e. $COR(c_0)$ corresponds to zero pigment concentration (setting $\alpha=0$) and depends on the $\eta_{LED}^{AM} / \eta_{LED}^{FM}$ ratio of the specific LED. The $COR(c_0)$ for D1 and D2 at $I_{LED}$=4 mA were respectively 3.92 ±0.08, and 8.38 ±0.17 indicating that the enhancement in $\eta_{LED}$ of D2 as compared to D1 is higher in AM than in FM.

Substituting $h$=230±50 μm, we obtain $\alpha(c_i)$. For example, using D2, $\alpha(c_3)$ = 14.5±3.1 cm$^{-1}$ is obtained. Thus, for the reference (undiluted) solution, $\alpha(c_{ref}) = (c_{ref}/c_3).\alpha(c_3)$=66±14 cm$^{-1}$. To obtain a quantitative estimate of $c_{ref}$, we first compute the emission-specific molar absorption coefficient corresponding to our broadband AM Si LED as $\alpha_M^{LED} = \int \alpha_M(\lambda).\epsilon(\lambda) \, d\lambda \approx 1410$ cm$^{-1}$mol$^{-1}$L, where $\alpha_M(\lambda)$ is the monochromatic molar absorption coefficient of carmine (Fig. 2d), $\epsilon(\lambda)$ [nm$^{-1}$] is the normalized AM EL-spectral irradiance [18] and integration limits are from $\lambda$=400 nm till 600 nm. Note that $\alpha_M^{LED}$ quantifies the overlap between the AM EL spectrum and the molar absorption coefficient spectrum of the target specimen. Subsequently, we obtain $c_{ref} = \alpha(c_{ref}) / \alpha_M^{LED}$=0.047±0.01 mol L$^{-1}$. To validate the result, the absorbance through a standard quartz cuvette with a 10 mm optical path of a solution sample with a concentration of 0.1% relative to $c_{ref}$ was measured using a commercial PerkinElmer Lambda 40 spectrophotometer at $\lambda$=525 nm. We obtained $\alpha_{525nm}(c_{ref})$ = 174 cm$^{-1}$, and hence $c_{ref}$ = $\alpha_{525nm}(c_{ref}) / \alpha_M(525\text{ nm})$ = 0.030 mol L$^{-1}$, in close agreement with that obtained using Si CMOS LED. The positive deviation in $c_{ref}$ as compared to the standard spectrophotometer technique is likely due to the underestimation in $\alpha_M^{LED}$ which can be caused by two factors. Firstly we ignored the weak absorption tail of carmine for $\lambda$ > 600 nm. Secondly, we ignored the presence of small amounts of curcumin, exhibiting a peak molar absorption coefficient at $\lambda$=425 nm [34] that can increase the absorption at the UV-edge of our spectrum: 400 nm < $\lambda$<450 nm. A maximum sensitivity of $S(c) \approx 1264$ cm$^{-1}$mol$^{-1}$L was calculated for D2 referred to concentration $c_3$ where $S(c)$ is defined as $\Delta\alpha/\Delta c$. The back-end oxide layer in our CMOS chip provided sufficient passivation to shield the Si LED from the chemicals in the droplet. Once a droplet was removed by cleaning with iso-propanol, the chip surface was re-used for the next droplet with negligible change in $\Delta I_{PD}$ measured in air. Concentrated glycerol exhibits boiling points exceeding 400 K and freezing points below 270 K [35]. Hence it serves as a non-volatile solvent for micro-volume sensing for a wide range of temperatures.

Our results form the basis of the very first experimental proof-of-concept small-volume (< 1 μL) optical absorption sensor with broad-spectrum avalanche-mode Si LEDs in standard CMOS technology. This obviates the need for expensive hybrid optical sources in applications. The color-ratio technique is expected to be immune to fluctuations in ambient temperature, as long as the thermal coefficients of LED quantum efficiency and that of the droplet-refractive index do not differ significantly between the optical wavelengths in forward and avalanche-mode EL spectra. Our sensor performance can be further improved by having a precise control over the volume and position of the droplet. Moreover, the placement of the Al-capped bond-pads can be optimized in order to avoid high electric fields within the droplet, which otherwise can affect the device reliability due to unwanted (electro-)chemical reactions between metals and glycerol [36][37] and other common polar solvents.

## IV. CONCLUSION

We reported the very first proof-of-principle of optical absorption sensing of pigment in solution with broad-spectrum silicon micro LEDs designed in a standard silicon-on-insulator CMOS technology. Vertical optical transmission through a glycerol micro-droplet containing carmine pigment was measured with a silicon PD while driving the on-chip silicon LED in both forward and avalanche modes of operation, and thereby steering it's electroluminescent spectrum between visible and near infrared. Hence, the same droplet can be used to measure optical transmission for both visible and near-infrared light from the same LED. The effective absorbance of the solution was obtained from the ratio of photocurrent in avalanche-mode to that in forward-mode LED operation. Further, from the observed droplet height and the known molar absorption coefficient of carmine, the effective absorption coefficient and carmine concentration was determined and validated by a commercial spectrophotometer.

## ACKNOWLEDGMENT

The work has been carried out under the Plantenna research program funded by the 4TU Federation of the Netherlands. The authors would like to thank NXP Semiconductors B.V. for silicon device fabrication, Gideon Emmaneel, Rob Luttjeboer, Agnieszka Kooijman and Nicky Dusoswa from TU Delft for technical support.

## REFERENCES

[1] R.A. Soref, "Silicon-based optoelectronics," proceedings of the IEEE, vol. 81, no. 12, pp. 1687–1706, December 1993.

[2] R. Soref, "Applications of silicon-based optoelectronics," MRS Bulletin, vol. 23, no. 4, pp. 20 – 24, April 1998.

[3] L.W. Snyman, H. Aharoni, A. Biber, A. Bogalecki, L. Canning, M. du Plessis, and P. Maree, "Optical sources, integrated optical detectors, and optical waveguides in standard silicon CMOS integrated circuitry," proceedings SPIE, vo. 3953, Silicon-based Optoelectronics II, pp. 20 – 36, 2000.

[4] U. Lu et al., "CMOS chip as luminescent sensor for biochemical reactions," IEEE Sensors J., vol. 3, no. 3, pp. 310 – 316 , June 2003.

[5] F.J. Blanco, M. Agirregabiria, J. Berganzo, K. Mayora, J. Elizalde, A. Calle, C. Domniguez, and L.M. Lechuga, "Microfluidic-optical integrated CMOS compatible devices for label-free biochemical sensing," J. Micromech. Microeng., vol 16, pp. 1006 – 1016 , April 2006.

[6] K. De Vos, I. Bartolozzi, E. Schacht, P. Bienstman, and R. Baets, "Silicon-on-insulator microring resonator for sensitive and label-free biosensing," Opt. Express, vol. 15, no. 12, pp. 7610 – 7615, June 2007.

[7] A. Frey, M. Schienle, and H. Seidel, "CMOS based sensors for biochemical analysis," proceedings IEEE Transducers, pp. 1670 – 1673, June 2009.

[8] C. Sun et al., "Single-chip microprocessor that communicates directly using light," Nature, vol. 528, pp. 534 – 538 , December 2015.

[9] A.H. Atabki et al., "Integrating photonics with silicon nanoelectronics for the next-generation systems on a chip," Nature, vol. 556, pp.349 – 354, April 2018.

[10] K. Xu, Y. Chen, T.A. Okhai, and L.W. Snyman, "Micro optical sensors based on avalanching silicon light-emitting devices monolithically integrated on chips," Optical Mat. Express, vol. 9, no. 10, pp. 3985 – 3997, October 2019.

[11] M. O' Toole and D. Diamond, "Absorbance based light emitting diode optical sensors and sensing devices," Sensors, vol. 8, pp. 2453 – 2479, April 2008.

[12] M. Green, J. Zhao, A. Wang, P.J. Reece, and M. Gal, "Efficient silicon light emitting diodes," Nature, vol. 412, no. 6849, pp. 805 – 808, 2001.

[13] T. Trupke, J. Zhao, A. Wang, R. Corkish, and M. Green, "Very efficient light emission from bulk crystalline silicon," Appl. Phys. Lett., vol. 82, no. 18, pp. 2996 – 2998, 2003.

[14] A.G. Chynoweth and K.G. McKay, "Photon emission from avalanche breakdown in silicon," Phys. Rev., vol. 102, no. 2, pp. 369 – 376, 1956.

[15] S. Dutta, R.J.E. Hueting, A.-J. Annema, L. Qi, L.K. Nanver, and J. Schmitz, "Opto-electronic modeling of light emission from avalanche-mode silicon p+n junctions," J. Appl. Phys., vol. 118, 114506, 2015.

[16] L.W. Snyman, M. du Plessis, and H. Aharoni, "Injection-avalanche-based n+pn silicon complementary metal-oxide-semiconductor light-emitting device (450-750 nm) with 2-order-of-magnitude increase in light emission intensity," Jpn. J. Appl. Phys., vol. 46, no. 4B, pp. 2474 – 2480, 2007.

[17] L.W. Snyman and K. Xu, "Stimulation of 450, 650, and 850 nm optical emissions from custom designed silicon LED devices by utilizing carrier energy and carrier momentum engineering," proceedings SPIE, vol. 10036, 1003603, 2017.

[18] S. Dutta, V. Agarwal, R.J.E. Hueting, J. Schmitz, and A.-J. Annema, "Monolithic optical link in silicon-on-insulator CMOS technology," Optics Express, vol.25, no. 5, pp. 5440 – 5456, March 2017.

[19] B. Huang, X. Zhang, W. Wang, Z. Dong, N. Guan, Z. Zhang, and H.Chen, "CMOS monolithic optoelectronic integrated circuit for on-chip optical interconnection," Opt. Communications, vol. 284, pp. 3924 – 3927, 2011.

[20] V. Agarwal, S. Dutta, A.-J. Annema, R.J.E. Hueting, J. Schmitz, M.J. Lee, E. Charbon, and B. Nauta, "Optocoupling in CMOS," proceedings IEEE 64th International Electron Devices Meeting, pp. 739 – 742, December 2018.

[21] P. Wessels, M. Swanenberg, H. van Zwol, B. Krabbenborg, H. Boezen,M. Berkhout, and A. Grakist, "Advanced BCD technology for automotive, audio and power applications," Solid-State Electronics, vol. 51, pp. 195 – 211, 2007.

[22] L. Gabirelli, D. Origgi, G. Zampella, L. Bertini, S. Bonetti, G. Vaccaro, F. Meinardi, R. Simonutti, and L. Cipolla, "Towards hydrophobic carminic acid derivatives and their incorporation in polyacrylates," R. Soc. Open Sci., vol. 5, 172399, June 2018.

[23] H. Croft and J. M. Chen, "Remote Sensing of Leaf Pigments," (Chapter) Comprehensive Remote Sensing, Shunlin Liang (ed.), Elsevier, Oxford, pp. 117-142, 2018.

[24] E.R. Hunt Jr., B.N. Rock, P.S. Rock, "Measuring of leaf water content by infrared reflectance," Remote Sensing of Environment, vol. 22, pp. 429 – 435, 1987.

[25] H.W. Gausman et al., "Refractive index of plant cell walls," Appl. Opt., vol. 13, no. 1, pp. 109 – 111, 1974.

[26] A. Bricaud, A. Morel and L. Prieur, "Absorption by dissolved organic matter of the sea (yellow substance) in the UV and visible domains," Limnology and Oceanography, vol. 26, no. 1, pp. 43 – 53, 1981.

[27] G. Pérez, C. Queimaliños, E. Balseiro, and B. Modenutti, "Phytoplankton absorption spectra along the water column in deep North Patagonian Andean lakes (Argentina)," Limnologica, vol. 37, pp. 3 – 16, 2007.

[28] Z.P. Lee and K.L. Carder, "Absorption spectrum of phytoplankton pigments derived from hyperspectral remote-sensing reflectance," Remote Sensing of Environment, vol. 89, pp. 361 – 368, 2004.

[29] M. Nitzan, A. Romem, and R. Koppel, "Pulse oximetry: fundamentals and technology update," Medical devices: evidences and research, pp. 231 – 239, July 2014.

[30] S. Dutta, P.G. Steeneken, V. Agarwal, J.Schmitz, A.J. Annema, and R.J.E. Hueting, "The avalanche-mode superjunction LED," IEEE Trans. Electron Devices, vol. 64, no. 4, pp. 1612 – 1618, April 2017.

[31] F. Behroozi and P.S. Behroozi, "Reliable determination of contact angle from the height and volume of sessile drops," Am. J. Phys., vol. 87, no. 1, pp. 28 – 32, January 2019.

[32] A. Moldovan, M. Bota, D. Dorobantu, I. Boerasu, D. Bojin, D. Buzatu, and M. Enachescu, "Wetting properties of glycerol on silicon, native SiO2, and bulk SiO2 by scanning polarization force microscopy," J. Adhesion Sci. and Tech., vol. 28, no. 13, pp. 1277 – 1287, March 2014.

[33] H. Kanemaru, S. Yukita, H. Namiki, Y. Nosaka, T. Kobayashi, and E. Tokunaga, "Giant Pockels effect of polar organic solvents and water in the electrical double layer on a transparent electrode," RSC Adv., vol. 7, pp. 45682 – 45690, 2017.

[34] K.I. Priyadarsini, "The chemistry of curcumin: from extraction to therapeutic agent," Molecules, vol. 19, pp. 20091 – 20112, December 2014.

[35] L.B. Lane, "Freezing points of glycerol and its aqueous solutions," Ind. Eng. Chem., vol. 17, no. 9, pp. 924, 1925.

[36] H.B. Zhang, J.H. Yang, F. Gao, and J.J. Lin, "Experimental study of the breakdown characteristic of glycerol as energy storage medium in pulse forming line," proceedings Annual Report Conference on Electrical Insulation and Dielectric Phenomena, pp. 850 – 853, 2013.

[37] S. Schünemann, F. Schüth, and H. Tüysüz, "Selective glycerol oxidation over ordered mesoporous copper aluminium oxide catalysts," Catal. Sci. Technol., vol. 7, pp. 5614 – 5624, 2017.